\documentclass[aps,pre,twocolumn,superscriptaddress,showpacs]{revtex4}

\usepackage{amssymb}
\usepackage{epsfig}
\usepackage{amsmath}
\usepackage{times}

\begin{document}

\title{The multiple effects of gradient coupling on network synchronization}
\author{Xingang Wang}
\affiliation{Temasek Laboratories, National University of Singapore, 117508, Singapore}
\affiliation{Beijing-Hong Kong-Singapore Joint Centre for Nonlinear \& Complex Systems
(Singapore), National University of Singapore, Kent Ridge, 119260, Singapore}

\author{Ying-Cheng Lai}
\affiliation{Department of Electrical Engineering, Department of
Physics and Astronomy, Arizona State University, Tempe, Arizona
85287, USA}

\author{Cangtao Zhou}
\affiliation{Institute of Applied Physics and Computational
Mathematics, P. O. Box 8009, Beijing 100088, P. R. China}

\author{Choy Heng Lai}
\affiliation{Department of Physics, National University of
Singapore, 117542, Singapore} \affiliation{Beijing-Hong
Kong-Singapore Joint Centre for Nonlinear \& Complex Systems
(Singapore), National University of Singapore, Kent Ridge, 119260,
Singapore}

\begin{abstract}
Recent studies have shown that synchronizability of complex
networks can be significantly improved by asymmetric couplings,
and increase of coupling gradient is always in favor of network
synchronization. Here we argue and demonstrate that, for typical
complex networks, there usually exists an optimal coupling
gradient under which the maximum network synchronizability is
achieved. After this optimal value, increase of coupling gradient
could deteriorate synchronization. We attribute the suppression of
network synchronization at large gradient to the phenomenon of
network breaking, and find that, in comparing with sparsely
connected homogeneous networks, densely connected heterogeneous
networks have the superiority of adopting large gradient. The
findings are supported by indirect simulations of eigenvalue
analysis and direct simulations of coupled nonidentical oscillator
networks.
\end{abstract}

\date{\today }
\pacs{05.45. Xt, 89.75.-k}
\maketitle

Complex networks have attracted a great deal of interest since the
discoveries of the small-world \cite{WS:1998} and scale-free \cite{BA:1999}
properties. Roughly, small-world networks are characterized by a locally
highly regular connecting structure and a globally small network distance,
while the defining characteristic of scale-free networks is a power-law
distribution $p\left( k\right) \sim k^{-\gamma }$ in the node degree.
Signatures of small-world and scale-free networks have been discovered in
many natural and man-made systems \cite{Strogatz:2001,AB:2002,Newman:2003},
and they constitute the cornerstones of modern network science.

At a systems level, synchronization is one of the most common
dynamical processes. For instance, in biology, synchronization of
oscillator networks is fundamental \cite{Strogatz:book}. In a
computer network designed for large scale, parallel computation,
to achieve synchronous timing is essential. Recent studies of the
synchronizability of complex networks have revealed that
small-world and scale-free networks, due to their small network
distances, are generally more synchronizable than regular networks
\cite{WC:2002,BP:2002}. A somewhat surprising finding is that a
scale-free network, while having smaller network distances than a
small-world network of the same size, is actually more difficult
to synchronize \cite{NMLH:2003}. Considering the ubiquity of
scale-free networks and the importance of synchronization in
network functions, the finding seems to have generated a paradox.
However, it is recently found that, with {\em weighted and
asymmetric couplings}, the synchronizability of scale-free
networks can be significantly improved and, in general, can be
much higher than the small-world networks
\cite{MZK:2005,HCAB:2005,CHAHB:2005,NM:2006,MZK:AIP,ZZWOR:2006,WLL:2007}.

For a pair of connected nodes on the network, the mutual couplings
between them are usually unbalanced. One direction will weight
over the other direction and generate a coupling gradient on the
link. To enhance network synchronization, both the direction and
the weight of coupling gradient should be properly set according
to the network properties such as node degree
\cite{MZK:2005,HCAB:2005} and betweenness \cite{CHAHB:2005}. In
Refs. \cite{MZK:2005,HCAB:2005} it has been shown that, by setting
the coupling gradient flow from the higher-degree node to the
smaller-degree node, the synchronizability of scale-free networks
can be significantly improved and higher than that of homogeneous
networks. With the same scheme of gradient direction, in Refs.
\cite{HCAB:2005,WLL:2007} it has been shown that the
synchronizability of scale-free networks can be further improved
by increasing the gradient \emph{weight}, and larger gradient in
general assumes higher synchronizability. This enhancing role of
coupling gradient is further highlighted in Ref. \cite{NM:2006},
where nodes are proposed to be connected by only gradient
couplings, i.e. the one-way-coupling configuration.

While most of the previous studies are focusing on the enhancing
role of coupling gradient on synchronization, there are
accumulating evidences showing the opposite: \emph{large gradient
may also deteriorates synchronization}. For example, in Ref. \cite
{MZK:2005,ZZWOR:2006} it is observed that as gradient increases,
network synchronizability is firstly enhanced, gradually reaching
to its maximum at an optimal gradient; then, after this optimal
value, increase of gradient will suppress synchronization. The
similar phenomenon is also briefly reported in Ref.
\cite{HCAB:2005}, there it is found that the suppressing effect of
large gradient has a close relation to the network parameters.
Despite of these observations, a detail study on the multiple
effects of coupling gradient on network synchronization is still
absent.

In this paper, we will study in detail the multiple effects of
coupling gradient on network synchronization, and investigate the
problem of synchronization optimization in asymmetrically coupled
scale-free networks. Our main findings are the following. (1) The
destructive role of large gradient comes from the phenomenon of
network breaking, increasing gradient will also increase the
breaking probability. (2) In general, small and densely-connected
heterogeneous networks have a lower breaking probability than
large and sparsely-connected homogeneous networks. (3) While large
gradient deteriorates the propensity of global synchronization,
partial synchronization of node clusters is enhanced.

We consider oscillator networks of the following form
\begin{equation}
\overset{.}{\mathbf{x}}_{i}=\mathbf{F}(\mathbf{x}_{i})-\varepsilon
\underset{j=1}{\overset{N}{\sum}}G_{i,j}\mathbf{H}(\mathbf{x}_{j}),\text{
}i=1,...N, \label{eq:model}
\end{equation}
where $\mathbf{F}(\mathbf{x}_{i})$ governs the local dynamics of
uncoupled node $i$, $\mathbf{H}(\mathbf{x})$ is the coupling
function, $\varepsilon $ is the coupling strength, and $G_{i,j}$
is an element of the coupling matrix $\mathbf{G}$ which takes form
\cite{MZK:AIP,ZZWOR:2006,WLL:2007}
\begin{equation}
G_{i,j}=-\frac{A_{i,j}k_{j}^{\beta }}{\sum_{j=1}^{N}A_{i,j}k_{j}^{\beta }},\
\ \mbox{for}\ \ i\neq j.  \label{eq:G_ij}
\end{equation}
with $k_{i}$ the degree of node $i$ and $A=\{a_{i,j}\}$ the adjacency matrix
of the network, $a_{i,j}=1$ if nodes $i$ and $j$ are connected, $a_{i,j}=0$
otherwise, and $a_{i,i}=0$. To keep the synchronization state a solution of
the system, we choose $G_{i,i}=1$.

It is worthy to note that the parameter $\beta $ in Eq.
(\ref{eq:G_ij}) modulates both the direction and weight of the
coupling gradient on each link \cite{WLL:2007}. Statistically, if
$\beta>0$, gradient is flowing from the higher-degree node to the
smaller-degree node; while if $\beta >0$, gradient is flowing in
the opposite direction. Meanwhile, by tuning the absolute value
$\left\vert \beta \right\vert$, we are also able to control the
gradient weight: larger $\left\vert \beta \right\vert $ generates
larger gradient \cite{WLL:2007}. Please also note that changing
$\beta$ does not change the total coupling cost of the network, it
only redistribute the weight of the couplings. Besides the
flexibility of gradient control, the coupling scheme of Eq.
({\ref{eq:G_ij}}) is also representative to a variety of network
models proposed in previous studies. For instance, the symmetric
network model in Refs. \cite{WC:2002,NMLH:2003} can be realized by
replacing $G$ with $A$ in Eq. (\ref{eq:model}); the weighted
asymmetric network model (constructed based on the information of
node degree) in Ref. \cite{MZK:2005} can be realized by setting
$\beta =0$ in Eq. (\ref{eq:G_ij}); and the directed tree-structure
network model in Ref. \cite{NM:2006} in principle can be realized
by setting $\beta \rightarrow \infty $ in Eq. (\ref{eq:G_ij}). A
schematic plot on the realization of these different network
models by changing $\beta $ is illustrated in Fig. 1.

\begin{figure}[tbp]
\begin{center}
\epsfig{figure=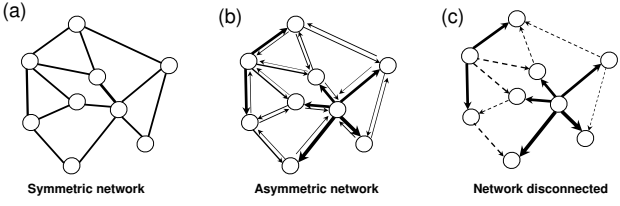,width=\linewidth}
\end{center}
\caption{A schematic plot on the transformation of a general
symmetrical network (a) to a weighted asymmetrical network (b) and
to a directed tree-structure network (the degenerative case)(c) by
increasing the gradient parameter $\protect\beta $ in Eq. (1) from
$0$ to $\infty$. The dashed lines in (c) denote the links which
are gradually diminished due to the increased gradient effect.}
\end{figure}

The limiting case of $\beta \rightarrow \infty $ in Eq.
(\ref{eq:G_ij}) is of special interest [Fig. 1(c)], since it
represents the extreme situation of node connection: the one-way
coupling format, i.e., $G_{i,j}=-1$ and $G_{j,i}=0$ when $k_j >
k_i$, and $G_{i,j}=0$ and $G_{j,i}=-1$ otherwise. In this case,
each node only receives coupling from one of its neighbors who has
the largest degree. (Strictly speaking, each node is receiving
coupling from one of its neighbors who has the largest node-scalar
$h_{i}=k_{i}^{\beta }\sum_{l\in V_{i}}k_{l}^{\beta }$, with
$V_{i}$ stands the set of neighboring nodes of $i$. However, for
complex networks of null degree-correlation, statistically we have
$h_{i}\sim k_{i}$. See Ref. \cite{WLL:2007} for details on the
definition of node scalar). For this one-way coupled network, if
it is non-degenerative (connective), nodes will be organized into
a unique tree-structure topology, with the largest-degree node of
the network locates at the root. Under this coupling
configuration, the network reaches its maximum synchronizability
\cite{NM:2006,WLL:2007}, manifesting the constructive role of
large gradient. However, if there are two or more large-degree
nodes on the network which are not directly connected, like the
case of Fig. 1, the network will be degenerative (disconnected)
[Fig. 1(c)]. In such a case, the network will break into several
subnetworks of tree-structure, and at the root of each subnetwork
locates a local largest-degree node. Once is broken, the network
can never be globally synchronized whatever the coupling strength,
manifesting the destructive role of large gradient.

To have a global picture on the gradient effects, we have
investigated the variation of network synchronizability as a
function of the gradient parameter $\beta$ for different network
topologies. The synchronizability of coupled network can be
evaluated by the method of master stability function (MSF)
\cite{BP:2002}, which states that a network is generally more
synchronizable when the spread of the eigenvalue spectrum of its
coupling matrix is narrow. In particular, let $0=\lambda _{1}\leq
\lambda _{2}\ldots \leq \lambda _{N}$ be the eigenvalue spectrum
of the coupling matrix $G$. Then the smaller the ratio $R\equiv
\lambda _{N}/\lambda _{2}$, the more likely synchronous dynamics
is to occur on the network. In general, the matrix constructed by
Eq. (\ref{eq:G_ij}) is asymmetric and its eigenvalues are complex.
Noticing that the coupling matrix $G$ can be written as
$G=QLD^{\beta }$, with $L=D-A$, $D=diag\{k_{1},k_{2},...k_{N}\}$
the diagonal matrix of degrees, and
$Q=diag\{1/\sum_{j}L_{1,j}k_{j}^{\beta},...1/\sum_{j}L_{N,j}k_{j}^{\beta}\}$
the normalization factors on rows of $G$. From the following
identity

\begin{equation}
\det (QLD^{\beta }-\lambda I)=\det (Q^{1/2}D^{\beta /2}LD^{\beta
/2}Q^{1/2}-\lambda I)
\end{equation}

it is found that the eigenvalues of the asymmetric matrix $G$ are equal to
that of the symmetric matrix $H=$ $Q^{1/2}D^{\beta /2}LD^{\beta /2}Q^{1/2}$,
which are real and nonnegative.

\begin{figure}[tbp]
\begin{center}
\epsfig{figure=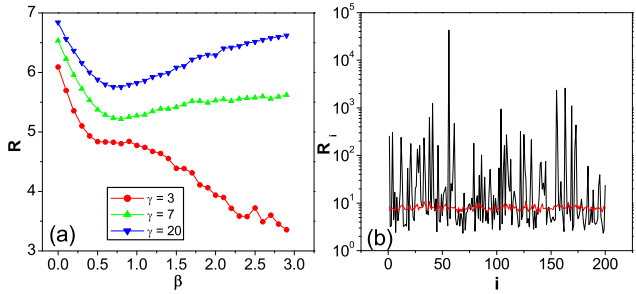,width=\linewidth}
\end{center}
\caption{(Color online) For scale-free networks of $N=2^{10}$ node
and average degree $\left\langle k\right\rangle =6$. (a) The
variation of eigenratio $R$ as a function of gradient parameter
$\protect\beta $ for networks of degree exponent $\protect\gamma
=3$ (the lower curve) and $\protect\gamma =7$ (the middle curve)
and $\protect\gamma =20$ (the upper curve). The optimal gradient
$\protect\beta _{o}$ is about $0.9$ in the later two cases. Each
data is an average result of $100$ network realizations. (b) For
the case of $\protect\gamma =7$ in (a), the value of $R_i$ for
different network realizations under parameters $\protect\beta
=0.5$ (the red curve) and $\protect\beta =5$ (the black curve).
The extreme values of $R_i$ indicate the broken of the network
topology in the corresponding realizations.}
\end{figure}

To simulate, we generate scale-free network of $N=2^{10}$ nodes
and average degree $\left\langle k\right\rangle =6$ by the
generalized model introduced in Ref. \cite{DM:2002}. In this
model, the degree exponent $\gamma $ can be adjusted via a
parameter $B$. Defining the new preferential attachment function
as $p\sim (k_{i}+B)/\sum_{j}(k_{j}+B)$, it can be proven that
$\gamma =3+B/m$, with $m=3$ the number of new links that
associated to each new added node in the model. Using $B=0, 12$
and $51$, we generate scale-free networks of degree exponents
$\gamma = 3, 7$ and $20$, respectively. The variations of $R$ as a
function of $\beta$ for these three networks are plotted in Fig.
2(a). It is found that, in the case of $\gamma=3$, increasing
$\beta$ will enhance synchronizability \emph{monotonically};
however, in the case of $\gamma=7$ or $8$, as $\beta$ increases
from zero, the network synchronizability is firstly enhanced, and,
after reaching its maximum at value about $\beta_{o} \approx 0.9$
\cite{COMMT:2}, it begin to be suppressed. Another interesting
finding in Fig. 2(a) is that, for the fixed gradient parameter
$\beta$, increasing the degree exponent $\gamma$ will
\emph{always} decrease the network synchronizability, indicating
the superior synchronizability of scale-free networks under
gradient couplings.

To gain insight on the transition of gradient effect from
enhancing to suppressing synchronization, we go on to investigate
the changes happening in the neighboring region of the optimal
gradient $\beta_{o}$. With $\gamma=7$ (the middle curve in Fig.
2(a)), we plot in Fig. 2(b) the individual value of $R_{i}$ for a
large number of network realizations under gradient parameters
$\beta _{1}=0.5<\beta _{o}$ and $\beta _{5}=5>\beta _{o}$. It is
found that, for $\beta _{1}=0.5$, the eigenratio $R_{i}$ is
oscillating around its mean value $R \approx 6$ with very small
fluctuations; while for $ \beta _{1}=5$, the eigenratio $R_{i}$
occasionally bursts into some extreme values of order $10^{3}$.
Since of $R \equiv \lambda _{N}/\lambda _{2}$, a divergent $R_{i}$
thus indicates $\lambda_2 \rightarrow 0$, which, from the
eigenvalue analysis, implies the breaking of network topology in
the corresponding realization. Therefore the suppression effect of
large gradient can be attributed to the phenomenon of network
breaking, and the optimal gradient can be understood as a balance
between the enhancing and the suppressing effect of gradient
coupling.

By knowing that synchronization suppression at large gradient is
induced by network breaking, we next to investigate the
relationship between the breaking probability and the network
parameters. To facilitate the analysis, we consider again the
limiting case of $\beta \rightarrow \infty$ in Eq.
(\ref{eq:G_ij}). As shown in Fig. 1(c), network breaking happens
when there are more than one local-maximum-degree nodes coexist on
the network. To break the network, the ``breaking nodes" do not
have to be possessing very large degree, they are only required to
have the largest degree among its neighbors. Once is broken, the
network will be divided into several tree-structured subnetworks,
with each subnetwork is led by one of such ``breaking nodes". Due
to the complicated configurations of the ``breaking nodes", we are
unable to give an analytical prediction on the relationship
between the breaking probability and the network parameters. The
numerical results on the variation of $p_{b}$ as functions of
$\gamma$ and $\left\langle k \right\rangle$ are plotted in Fig. 3,
which show that, in comparing sparsely connected homogeneous
networks, densely connected heterogeneous networks are more
difficult to break down \cite{COMMT:1}. It can be also found from
Fig. 3(b) that the increase of system size will always increase
the breaking probability. These findings (calculated for the
limiting case) are coincident with the findings in Fig. 2
(calculated for the general case), both tell us that heterogeneous
networks are more sustainable to large gradient.

\begin{figure}[tbp]
\begin{center}
\epsfig{figure=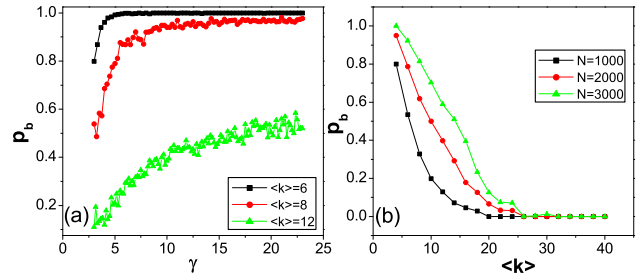,width=\linewidth}
\end{center}
\caption{(Color online) For scale-free network of $N=3\times
10^{3}$ nodes and under the limiting case of $\protect\beta
\rightarrow \infty$ in Eq. (\ref{eq:G_ij}), the probability of
network breaking $p_{b}$ as a function of the degree exponent
$\protect\gamma $ (a) and the average degree $\left\langle
k\right\rangle$ (b). Each data is an averaged result of $1000$
network realizations.}
\end{figure}

The results exemplified in Figs. 2 and 3 are based on the
eigenvalue and topology analysis. It is useful to examine the
gradient effects in real oscillator networks. For this purpose we
have check the synchronization of scale-free networks of coupled
nonidentical chaotic R\"{o}ssler oscillators, a typical model
employed in studying network synchronization \cite
{BP:2002,MZK:2005,CHAHB:2005,HCAB:2005}. The dynamics of a single
oscillator is described by
$\mathbf{F}_{i}(\mathbf{x}_{i})=[-\omega _{i}y_{i}-z_{i},\omega
_{i}x_{i}+0.15y_{i},z_{i}(x_{i}-8.5)+0.4]$, where $\omega _{i}$ is
the natural frequency of the $i$th oscillator. In simulations we
choose $\omega _{i}$ randomly from the range $[0.9,1.1]$, so as to
make the oscillators nonidentical. The coupling function is chosen
to be $\mathbf{H}(\mathbf{x})= \mathbf{x}$. The degree of
synchronization can be characterized by monitoring the amplitude
$A$ of the mean field $X(t)=\sum_{i=1}^{N}x_{i}(t)/N $
\cite{MZK:2005}. For small coupling strength $\varepsilon $,
$X(t)$ oscillates irregularly and $A$ is approximately zero,
indicating lack or a lower degree of synchronization. As the
coupling parameter is increased, synchronization sets in and $A$
is increased gradually from zero (nonsynchronous state) to its
maximum (synchronous state). By $\varepsilon =0.15$, we plot in
Fig. 4(a) the variation of $A$ as a function of $\beta $. It is
found that as $\beta $ increases $A$ is firstly increased, and
reaching its maximum at about $\beta _{o}\approx 1$, manifesting
the constructive role coupling gradient. Then, as $\beta$
increases from $\beta _{o}$, $A$ begins to decrease, manifesting
the destructive role of coupling gradient. An interesting finding
is that increasing $\beta$ further does not decreases $A$
continuously, the system always keeps on high coherence at about
$A\approx 8$. Please note that for the adopted coupling function
$\mathbf{H}$, the stable region of the MSF function \cite{BP:2002}
has only a lower boundary, which confirms that the decrease of $A$
at $\beta \geq \beta_o$ in Fig. 4(b) is induced \emph{exclusively}
by the breaking effect (not induced by the instability of the
shortest wave mode \cite {BP:2002}).

\begin{figure}[tbp]
\begin{center}
\epsfig{figure=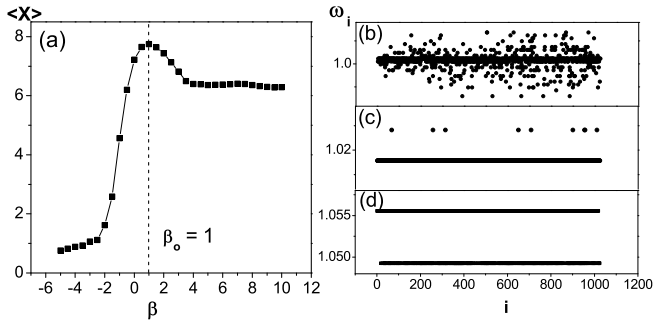,width=\linewidth}
\end{center}
\caption{For scale-free network of $N=2\times 10^{3}$ nonidentical
R\"{o} ssler oscillators and average degree $\left\langle
k\right\rangle =6$. (a) For $\varepsilon=0.15$, the variation of
the mean field amplitude $A$ as a function of the gradient
parameter $\beta$. Network synchronization is optimized at about
$\beta \approx 1$. Each data is an averaged result of $50$ network
realizations. For a chosen network realization of (a), the
frequency distribution of the oscillators under gradient
parameters $\beta=0.1$ (b), $\beta=1$ (c) and $\beta=10$ (d).}
\end{figure}

The finding that $A$ keeps on large values at very large $\beta $
indicates that, despite of the increased probability of network
breaking, nodes are still strongly correlated, with a manner that
is different to the situation of small $\beta $. To gain insight,
we choose a network realization in Fig. 4(a), and plot the
frequency distribution $\omega_i$ of the oscillators under
different gradient parameters: $\beta=0.1$ in Fig. 4(b), $\beta =
1$ in Fig. 4(c) and $\beta=10$ in Fig. 4(d). It is found that, for
small gradient [Fig. 4(b)], $\omega_i$ is distributed randomly
around the mean value $\omega =1$, indicating the low system
coherence; around the optimal gradient [Fig. 4(c)], most of the
oscillators are synchronized to be having the same frequency, with
few exceptions which are synchronized to another frequency,
indicating a higher coherence of the system dynamics under this
gradient; for large gradient [Fig. 4(d)], the oscillators are
separated into two clusters of the similar size, the frequency of
one cluster is different to that of the other one. Fig. 4(d)
indicates that, under the large gradient, although the probability
of network breaking is high, network nodes are still strongly
correlated due to the existence of synchronous clusters. The
clusters are a direct result of the network breaking. Led by the
``breaking nodes", each subnetwork develops into a synchronous
cluster.

In summary, we have studied the multiple effects of coupling
gradient on network synchronization, and investigated the
dependence of these effects to the network parameters. Our
findings suggest that, in comparing with sparsely connected
homogeneous networks, densely connected heterogeneous networks
take more advantages from the gradient couplings.

YCL and CTZ thank the great hospitality of National University of
Singapore, where part of the work was done during their visits.
YCL was supported by AFOSR under Grants No. FA9550-06-1-0024 and
No. FA9550-07-1-0045. CTZ was supported by the National Natural
Science Foundation of China under Grant No. 10575013.


\begin{thebibliography}{99}
\bibitem{COMMT:1} Breaking phenomenon is also found
in other types of complex networks such as the modular and
assortative networks, there breaking prabilities are also
dependent on the network modularity and assortativity.
\bibitem{COMMT:2} Numerically we find that $\beta_o$ is insensitive to $\gamma$.
This can be heristically understood as a competing result of the
breaking probability and the total gradient. Under the same
parameter $\beta$, heterogeneous networks have small breaking
probability but possess large gradient, the former enhances
synchrnization while the latter in general suppresses
synchronization.
\end{thebibliography}
\end{document}